\documentclass[aps,prl,floatfix,showpacs,twocolumn,tightenlines,superscriptaddress]{revtex4}

\usepackage{amsmath}
\usepackage{graphicx}
\usepackage{epsfig}

\begin{document}

\title{Guidance and Control in a Josephson Charge Qubit}

\author{J.\ F.\ Ralph}\email{jfralph@liv.ac.uk}
\affiliation{Department of Electrical Engineering and Electronics, The University of Liverpool, 
Brownlow Hill, Liverpool, L69 3GJ, United Kingdom}

\author{E.\ J.\ Griffith}
\affiliation{Department of Electrical Engineering and Electronics, The University of Liverpool, 
Brownlow Hill, Liverpool, L69 3GJ, United Kingdom}

\author{T.\ D.\ Clark}
\affiliation{School of Engineering, The University of Sussex, 
Falmer, Brighton, BN1 9QT, United Kingdom. }

\author{M.\ J.\ Everitt}
\affiliation{School of Engineering, The University of Sussex, 
Falmer, Brighton, BN1 9QT, United Kingdom. }

\date{January 12, 2004}

\begin{abstract}
In this paper we propose a control strategy based on a classical
guidance law and consider its use for an example system: a Josephson charge qubit.
We demonstrate how the guidance law can be used to attain a desired qubit state
using the standard qubit control fields.
\end{abstract}

\pacs{03.65.-w, 74.50.+r, 85.25.Dq}

\maketitle

\section{Introduction}

\noindent
The practical operation of devices for quantum information processing is dependent upon the
ability to control the behaviour of the component qubits via external classical control fields.
As in classical devices, the control/bias fields allow the operator to define the dynamical
characteristics of the system. Bias fields contain noise that will couple into the device
and can ultimately limit the coherent evolution of a quantum information processing system.
In classical systems, a feedback control loop is often used to reduce the effect of such 
environmental noise or other unforseen perturbations on the evolution of system. The control
loop compares the desired behaviour with the actual behaviour of the system and aims to minimise
the error between the two. The problem for quantum control is that a feedback loop 
requires some form of measurement to be made, and this measurement will often adversely affect
the coherence of the quantum evolution. Several groups have suggested methods to overcome this
problem, using techniques developed (mainly in quantum optics) to describe `weak' measurements. 
These measurements can be used to obtain information about a quantum system over a period of time whilst 
minimising the adverse effects of the measurement interaction (see reference \cite{Doh00a} for a
recent review and a description of the relationship between classical control and quantum control).
Closed-loop techniques fall into two main categories: Markovian feedback 
\cite{Wis94,Hof98,Wan01} and Bayesian (or optimal) feedback \cite{Doh00a,Bel99,Doh00b}. The first 
method uses the results of measurements to directly alter the external control fields applied to 
the system. The second builds an estimate of the system state over a number of measurements. 
Although they were developed in quantum optics, these techniques and related analysis have recently 
been applied to the control of solid-state qubits \cite{Kor99,Oxt04}. 

This paper deals with an associated problem, that of generalising the techniques of classical
guidance (see for example \cite{Zar97}) to the operation and manipulation of qubits. The main
conceptual difference between guidance and control is one of timeliness. In control systems, the 
desired state of the system (classical or quantum) may be static or change with time, but there is 
always an error between the actual state of the system and the desired state. A 
control is applied to remove this error signal. In guidance systems, the evolution
of the system is not as important as the final state. The controls are applied throughout the evolution
to ensure that the system reaches the desired state at the desired time. In \cite{Bou04} Bouten et al. 
have addressed this problem implicitly, by using dynamical programming to solve an optimal control 
problem by minimising the controls (which define a `cost' function) applied over the time available ($t_{max}$). 
Experience with classical guidance techniques shows that, whilst such algorithms may give a 
minimum cost solution, optimal control guidance can be difficult to implement and simpler guidance
laws often provide sufficient accuracy with significantly simpler guidance-control systems \cite{Zar97}.
The most commonly used guidance law is referred to as {\it proportional navigation}, which is used
in a wide variety of aerospace guidance systems (autopilots, guided missiles, etc.). Several
variants of proportional navigation exist, but - in its the most general form - it can be written
as \cite{Zar97},
$$
a_c = N'\frac{(ZEM)}{t_{go}^2}
$$
where $a_c$ is the control (acceleration command) that should be applied to the system, $N'$ is a constant
(called the `navigation constant') which determines the strength of the commands, $t_{go}$ is the
time to go until the objective ($t_{max}\ge t_{go} \ge 0$), and $ZEM$ is the `zero effort miss' 
(that is, the distance between the
desired state - the intercept point - and the predicted state if no more controls are applied). In classical 
guidance, an intercept is assured as long as $N'>2$ and the accelerations commanded are achievable. In 
practice, $N'$ is normally in the range $N' = 4 \rightarrow 6$, so that the controls immediately prior
to intercept are minimised.

Proportional navigation guidance is not optimal in the sense of minimising the controls applied, but  is
generally easier to implement in a practical control system and the controls that need to be applied to the 
system tend to have a lower bandwidth than those generated by more sophisticated algorithms.
In this paper, we will generalise this classical guidance law to the problem of controlling a solid-state
qubit. We will show some examples of the behaviour predicted for an (open-loop) proportional guidance law 
applied to a Josephson charge qubit (e.g. \cite{Nak99,Nak03}). (Open-loop control, i.e. 
without feedback, has been studied in the context of atomic and quantum optics but not 
in the same form as that presented here \cite{Vio99}). In particular, we consider the robustness of the
guidance law to noise in the bias fields and the introduction
of a first-order time delay (low pass filter) into the control system, to investigate the effect of
restricting the bandwidth of the control signal. Consideration is also given to the introduction
of a simple measurement interaction and feedback control loop.

\section{Josephson Charge Qubit}

\noindent
The qubit studied in this paper is a commonly used (idealised) model for a standard experimental configuration. 
It consists of a superconducting island (a Cooper pair box) coupled to an external circuit via two parallel 
Josephson junctions \cite{Nak99,Nak03} (see Figure 1). The 
qubit has two main control fields, a voltage bias ($V_x$) to control the energy of the charge states, and a 
magnetic flux ($\Phi_x$) to control the tunnelling of electron pairs between the box and the external circuit. The 
two parallel Josephson junctions form a current loop and applying a magnetic field through this loop 
acts so as to modulate the tunnelling rate onto and off the Cooper pair box. (In this paper, we assume 
that the two Josephson junctions are identical for simplicity. In practice, there will be small variations 
in the tunnelling rates for each junction in any experimental system and it might be necessary to characterise 
these differences in a real system). The effect of modulating the effective tunnelling frequency on the 
qubit energy levels is shown in Fig1(b). We shall use circuit 
parameter values based on the experimental values given in \cite{Nak99,Nak03} to ensure that the circuit
parameters are realiseable. 
In most experiments that have been reported using such systems (as well as in 
other superconducting qubit experiments based on persistent current devices \cite{Fri00}), excited states 
are generated in the qubit by applying an additional field, a time-dependent microwave drive field. In this 
paper, we do not use an additional (external) microwave drive which reduces the complexity of the control system. 
This point it discussed in more detail below.

\begin{figure}
\begin{center}
\includegraphics[height=12cm]{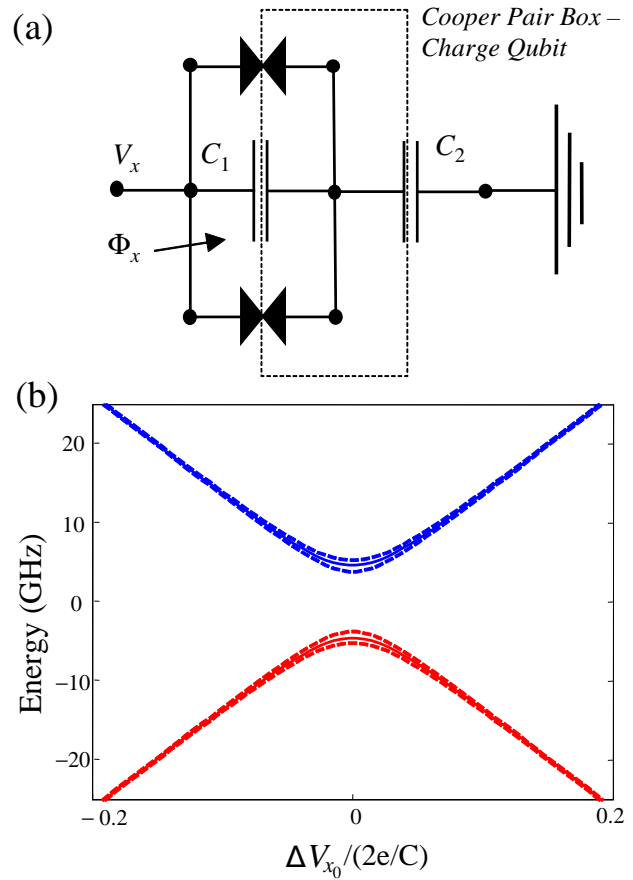}
\end{center}
\caption{(Color online) (a) Schematic diagram of qubit circuit, showing capacitances $C_1+C_2=C$; (b) Energy levels for
qubit as a function of $V_x$, using parameters given in the text and showing energies for nominal bias
point $\Phi_{x_0} = 0.25 \Phi_0$ (solid lines) and for extremes of the control modulation field 
$\Phi_{x_0} = 0.25\pm 0.05 \Phi_0$(dashed lines).}
\end{figure}

The Hamiltonian for the (two-state) qubit can be written in the charge basis representation as 
\cite{Nak99,Nak03},
\begin{equation}
H_0 = \left(
\begin{array}{cc}
\frac{CV_x^2}{2} & -\frac{\hbar\nu}{2}\cos\left(\frac{\pi\Phi_x}{\Phi_0}\right) \\
-\frac{\hbar\nu}{2}\cos\left(\frac{\pi\Phi_x}{\Phi_0}\right) & \frac{(2e-CV_x)^2}{2C}
\end{array}\right)
\end{equation}
where the basis states are zero excess pairs ($|0\rangle$) and one excess
pair ($|1\rangle$) on the Cooper pair box, $C$ is the net capacitance of the island/box
(in this case, we take $C=6\times 10^{-16}$ F), and $\nu$ is the tunnelling (angular)
frequency of the Josephson junction. The maximum Josephson tunnelling frequency is taken to be
$\nu/2\pi \simeq 12.9$ GHz, although the effective Josephson frequency at the nominal
bias point $\Phi_x=\Phi_0/4$ is approximately 9.1 GHz (see below), in line with the 
parameters given in \cite{Nak03}, where $\Phi_0=h/2e=2\times 10^{-15}$ Wb is the superconducting 
flux quantum. For the purposes of this paper, we take the charge basis to be the computational
basis for the qubit. The energy eigenstates are functions of the bias fields. Although we
we will consider transitions between energy eigenstates under the action of the guidance/control, 
this is not necessary. As discussed below, we use the ground state as the initial state for
convenience, since it is assumed that the system will relax to this state after some period
of time, under the action of whatever dissipation processes are present in the system.

This Hamiltonian (or any other $2\times 2$ Hermitian matrix) may be decomposed into four components, 
which correspond to a constant multiplied by the identity matrix ($I$) or one of the three Pauli
matrices 
$$\sigma_x = \left(\begin{array}{cc} 0 & 1 \\ 1 & 0 \end{array}\right)
,\hspace{3mm} \sigma_y = \left(\begin{array}{cc} 0 & -i \\ i & 0 \end{array}\right)
,\hspace{3mm} \sigma_z = \left(\begin{array}{cc} 1 & 0 \\ 0 & -1 \end{array}\right)
$$ 
We write the bias voltage as $V_x=e/C+\Delta V_x$, and then
decompose the Hamiltonian as,
\begin{eqnarray}
H_0 &=& H_I I+H_x \sigma_x+H_y \sigma_y+H_z \sigma_z \nonumber \\
&=&\left(\frac{C(\Delta V_x)^2}{2}+\frac{e^2}{2C}\right)I+e(\Delta V_x) \sigma_z \nonumber \\ &&-\frac{\hbar\nu}{2}\cos\left(\frac{\pi\Phi_x}{\Phi_0}\right) \sigma_x
\end{eqnarray}
From this decomposition, it is easy see that there is no $\sigma_y$ term in the basic Hamiltonian.
This would be the term that would normally be responsible for exciting the qubit into the excited state
and has been the most common control coupling to be studied in quantum optics \cite{Wan01,Doh00b,Bou04,Wis02}.
To generate a $\sigma_y$ term, it is necessary to apply a time-dependent field, since $\sigma_x$ and $\sigma_z$
do not commute: $[\sigma_z, \sigma_x] = 2i\sigma_y$. Normally, in quantum optics, a laser is used to pump a
qubit into an excited state, or in solid-state experiments (such as those described in references \cite{Nak99}
and \cite{Nak03}) an external microwave source is used. In an experimental system, the underlying Hamiltonian
may not be exactly what is predicted by the idealised model used here, but a number of techniques have been
proposed to allow the deviations to be characterised \cite{Ral03}. The use of an external microwave source in solid-state is not ideal for large scale systems because of potential problems in isolating qubits from drives
applied to neighbouring devices. Because of this, we restrict ourselves to controls that arise from
time-dependent bias fields $\Delta V_x$ and $\Phi_x$ and consider the effect of limiting the bandwidth 
of these fields in a later section.

The general representation for a qubit state is,
\begin{equation}
|\psi\rangle = \cos\left(\frac{\theta}{2}\right)|0\rangle+\sin\left(\frac{\theta}{2}\right)e^{i\phi}|1\rangle
\end{equation}
where $\theta \in [0,\pi]$ and $\phi \in [0,2\pi]$, which can be written as a (pure state) density 
matrix $\rho = |\psi\rangle \langle\psi |$,
\begin{equation}
\rho = \left(\begin{array}{cc} \cos^2\left(\frac{\theta}{2}\right) & 
\cos\left(\frac{\theta}{2}\right)\sin\left(\frac{\theta}{2}\right)e^{-i\phi} \\ 
\cos\left(\frac{\theta}{2}\right)\sin\left(\frac{\theta}{2}\right)e^{i\phi} & \sin^2\left(\frac{\theta}{2}\right) \end{array}\right)
\end{equation}
However, the most convenient representation for the purposes of this paper is the
Bloch sphere representation \cite{Scu97}, where the two angles $\theta$ and $\phi$
represent angles on a unit sphere, defined in a three-dimensional space by,
\begin{equation}
\left(\begin{array}{c} X \\ Y \\ Z \end{array}\right) = 
\left(\begin{array}{c} \sin\theta\cos\phi \\ 
\sin\theta\sin\phi \\
\cos\theta \end{array}\right) = 
\left(\begin{array}{c} \rho_{01}+\rho_{10} \\ i\rho_{01}-i\rho_{10} \\
\rho_{00}-\rho_{11} \end{array}\right)
\end{equation}
The different components ($\sigma_x$, $\sigma_y$ and $\sigma_z$) present in the Hamiltonian represent 
rotations in this three-dimensional Bloch space (about the $X$, $Y$ and $Z$ axes respectively). The fact 
that there is no $\sigma_y$ term in the basic Hamiltonian is not a problem, because it is possible to
reach any point on the Bloch sphere from any other by successive rotations about any two (non-parallel)
axes. The guidance algorithm simply governs the size of the rotations that are to be applied to achieve
the objective.

\section{Proportional Guidance}

\noindent
The classical proportional navigation algorithm predicts the expected miss distance if no control is applied
(the `Zero Effort Miss' or $ZEM$), a quantum analogue for the qubit can be developed in the similar manner. 
The evolution in the absence of controls is described by the basic Hamiltonian given in 
equation (2) and the time evolution of the wavefunction and (pure state) density matrix is governed by a 
unitary evolution operator,
$$
\hat{U}(t) = \exp\left(-\frac{i H_0 t}{\hbar}\right)
$$
The main difference between proportional navigation on a sphere and proportional navigation in 
three Euclidean dimensions is that the rotations generated by this unitary matrix and the rotations
required to move the estimated final state onto the desired state do not commute. The rotation 
required at the end point will not produce the same effect if it is applied earlier in the trajectory.
Because of this, we need to {\it retrodict} where the desired state should have been at the earlier time,
if it is to end up at the desired state under the free evolution given by $H_0$. The $ZEM$ in this case
is the two angles ($\theta_{ZEM}$ and $\phi_{ZEM}$) which separate the current estimated state (on the
Bloch sphere) from the point where the desired state would have to be at the current time. So, we calculate
\begin{eqnarray}
\rho_{d}(t_{go}) &=& \hat{U}^{\dagger}(t_{go})\rho_{d}\hat{U}(t_{go}) \nonumber \\
&=& \exp\left(\frac{i H_0 t_{go}}{\hbar}\right)\rho_{d}\exp\left(-\frac{i H_0 t_{go}}{\hbar}\right)
\end{eqnarray}
where $\rho_d$ is the desired final state (or `target state'). From this density matrix, we can calculate 
the two angles $\theta_{d}(t_{go})$ and $\phi_{d}(t_{go})$ which define the retrodicted state. The $ZEM$ angles
are then given by,
\begin{eqnarray}
\theta_{ZEM} &=& \theta_{d}(t_{go}) - \theta \nonumber \\
\phi_{ZEM} &=& \phi_{d}(t_{go}) - \phi
\end{eqnarray}
where $\theta$ and $\phi$ represent the current state (allowing for the periodicity of the angles). 
The controls that need to be applied are
\begin{eqnarray}
\frac{d\theta_c}{dt}&=&N'\frac{(\theta_{ZEM})}{t_{go}} \nonumber \\
\frac{d\phi_c}{dt}&=&N'\frac{(\phi_{ZEM})\sin\theta}{t_{go}} \nonumber \\
\end{eqnarray}
where the controls are angular velocities rather than accelerations because the Bloch equations are 
first-order differential equations \cite{Scu97}, rather than second-order classical dynamics, and 
the $\sin\theta$ term arises because the differences in $\phi_{ZEM}$ near the poles of the Bloch sphere
need to account for the curvature of the sphere.

The controls can be equated to an equivalent Hamiltonian by integrating over a small time interval
($\delta t$) and using the fact that a $\sigma_x$-Hamiltonian generates rotations about the $X$-axis
and a $\sigma_z$-Hamiltonian generates rotations about the $Z$-axis. We find the Hamiltonian that
rotates the Bloch vector from its current position (as given by $\theta$ and $\phi$) through angles, 
\begin{eqnarray}
\delta \theta_c&=&N'(\delta t)\frac{(\theta_{ZEM})}{t_{go}} \nonumber \\
\delta \phi_c&=&N'(\delta t)\frac{(\phi_{ZEM})\sin\theta}{t_{go}} 
\end{eqnarray}
Making a linear approximation and solving for the Hamiltonian controls ($H_{x_c}$ and $H_{z_c}$), we
obtain expressions,
\begin{eqnarray}
H_{x_c} &=& -\frac{\hbar(\delta\theta_{c})}{2(\delta t)\sin\phi} \nonumber \\
H_{z_c} &=& -\frac{\hbar(\delta\phi_{c})}{2(\delta t)}
\end{eqnarray}
(In practice, although the expression for $H_{x_c}$ includes a $1/\sin\phi$ term, removing the dependence
upon $\phi$ does not affect the performance of the guidance to a large degree and it dramatically
reduces the bandwidth required for the control signal and makes the control system more resilient to time
delays). 

Applying these controls requires manipulating the bias fields, $\Delta V_x$ and $\Phi_x$.
Clearly there are limits to the size of the controls that can be applied using these  bias fields. 
The two state approximation for the charge qubit is only valid as long as changes in the bias voltage  
are small $|\Delta V_x| \ll 2e/C$, so we impose constraints, such that $|\Delta V_{x_c}| < 0.1\times 2e/C$ where
$\Delta V_{x_0}$ is the nominal voltage bias point and the voltage bias control fluctuates 
about this point $\Delta V_{x} = \Delta V_{x_0}+\Delta V_{x_c}$. Here the capacitance that is used to apply the
gate voltage, which in turn controls the voltage applied across the qubit, is assumed to be the same as the 
qubit island capacitance. This is not necessary, changing the gate capacitance simply rescales the voltage bias
and the associated behaviour under the action of the guidance/control. As long as this capacitance is known, 
the appropriate controls can be applied.
For the magnetic flux bias, we must ensure that any fluctuation around the nominal bias point $\Phi_{x_0}$
is small enough so that the response is approximately linear. So $\Phi_x =
\Phi_{x_0}+\Delta\Phi_{x_c}$ where $\Phi_{x_0}=0.25 \Phi_0$ and $|\Delta\Phi_{x_c}|<0.05 \Phi_{0}$. Within 
this region, the cosine tunnelling term is approximately linear in $\Delta\Phi_{x_c}$ and we obtain the
following relations for the control fields,
\begin{eqnarray}
\Delta\Phi_{x_c} &=& \frac{2 H_{x_c}}{\pi\hbar\nu\sin\left(\frac{\pi\Phi_{x_0}}{\Phi_0}\right)} \nonumber\\
\Delta V_{x_c} &=& \frac{H_{z_c}}{e}
\end{eqnarray}
Beyond the limits given, the control is assumed to have saturated and any controls commanded are unachievable.
This limits the number of states that are reachable from a given initial state, but as long as the desired state
falls within the reachable set for the initial state, this should not be a major problem.
For simplicity, we assume that the initial state corresponds to the ground state for the unperturbed
Hamiltonian $H_0$ at the nominal bias point ($\Delta V_{x_0}$ and $\Phi_{x_0}=0.25 \Phi_0$). That is, for simplicity,
we assume that the qubit has relaxed into the ground state via some (weak) dissipative process.
Although it would be possible to prepare a different initial state by another process, such as some type of
projective measurement interaction, this is not considered here. 

The guidance algorithm operates by integrating the evolution of the qubit for a small timestep 
and applying controls via $\Delta V_x$ and $\Phi_x$ that are determined by finding the $ZEM$ angles
from equations (6) and (7), converting these angles into an effective Hamiltonian using equations (9)
and (10), and finally converting these Hamiltonian controls to bias values via equation (11). Applying 
this procedure iteratively generates a time-dependent bias signal that can then be applied to a qubit and 
should provide the desired state at the desired time ($t_{go}=0$) as long as it is within the 
reachable set. Of course, since the control is currently open-loop, the actual state of the system is unknown.
The controls are generated from the knowledge of where the state should be, if it started in the
ground state and the controls had been correctly applied. A wide variety 
of states may be prepared in this way and the controls have several distinct advantages: they have
a comparatively low bandwidth, they operate via the standard bias fields without an additional external drive,
and (for states within the reachable set) the controls vanish as $t_{go}\rightarrow 0$. This last point means that
the Hamiltonian is only weakly perturbed at the time when the state is required. Although the algorithm 
is presented as an open-loop control technique, it provides a natural generalisation to feedback (closed-loop) 
control, which is discussed in a later section.

The difference between the desired state and the final (possibly mixed) state can be quantified
in a variety of standard ways. For the purposes of this paper, we use three common measures
to characterise the performance of the guidance algorithm. The first 
measure is the {\it fidelity} of the state, introduced by Jozsa \cite{Joz94}, which is one measure for how 
close the final state is from the desired state. The fidelity $F$ for two density matrices, 
$\rho_{f}$ and $\rho_{d}$ is given by \cite{Pet04},
$$
F = F(\rho_{f},\rho_{d}) = \left|{\rm Tr}\left[\sqrt{\sqrt{\rho_{d}} \rho_{f} \sqrt{\rho_{d}}}\right]\right|^2
$$
which varies between zero and one (one being that the final state matches the desired state exactly).
The second measure that we use in this paper is the {\it trace distance} $D$ \cite{Nie00},
$$
D = D(\rho_{f},\rho_{d}) = \frac{1}{2}{\rm Tr}[\rho_{f}-\rho_{d}]
$$
which also runs from zero to one (zero being identical states) and measures the separation
of the desired and final states. The final measure is the distance from the surface
of the Bloch sphere, which is a measure of the purity (or, conversely, the mixedness) of the state, 
and can be written as \cite{Wan01,Wis02},
$$
p = 2{\rm Tr}[\rho^2]-1
$$
with pure states giving $p = 1$ and maximally mixed states giving $p = 0$. The purity is most
important for mixed states, which are generated by the inclusion of stochastic terms, such as the
measurement and feedback model discussed in a later section.

\section{Ideal Open-Loop Performance}

\noindent
Starting the qubit in its ground state, we can pick a particular target state and calculate the 
control signal required to generate this state at a later time using the technique described above.
We start by examining the performance of the guidance algorithm when applied to an ideal system
(with no errors) and then examine the performance in the presence of possible experimental errors. 

Although any target state can be chosen, an obvious candidate is the qubit excited state (which is
a function of the nominal bias fields). As we have said, we assume that the initial qubit state will
be the ground state, the qubit having relaxed into this state prior to the control being applied.
Figure 2(a) shows the fidelity and trace distance achieved for a range of $\Delta V_{x_0}$ and for three
different time periods $t_{max}= 20, 50, 100$ qubit cycles and $N' = 8$. The purity is trivially one, since there 
is no stochastic evolution and the qubit state is therefore pure. We see from the graphs that the 
excited state is easier to reach (high fidelity and low trace distance) from voltage values 
close to zero, $\Delta V_{x_0}\simeq 0$. The fidelity
is one for a wide range of bias voltages, even when the time available is relatively short, 
$t_{max}= 20$ cycles. The main reason that the excited state is easier to reach near zero bias is that
the ground and excited states at zero bias are equal (symmetric and anti-symmetric) superpositions of the
two basis states, and the main controls that are required are rotations about the Z-axis ($\sigma_z$), which
are easier to achieve than rotations about the X-axis or Y-axis. As the bias voltage is increased, the
ground and excited states shift toward the poles of the Bloch sphere (and the qubit natural oscillation
frequency increases, reducing the time available), and more controls are required from the magnetic
bias field (which generates $\sigma_x$ terms). However, allowing more time for the guidance (by extending
$t_{max}$) increases the ability to generate an excited state, and for $t_{max} = 100$ cycles
the excited state is within the reachable set for 
the whole of the range $\Delta V_{x_0}$ shown in Figure 2(a).
Selecting the excited state as a target state has one more practical advantage. Because the 
exited state is  stationary by definition, the production of an excited state is quite robust.
Slight variations in the control fields or the time to go only generate small deviations from
the excited state. 

Figure 2(b) shows the performance of the guidance algorithm for a target state that is not an
energy eigenstate for any values of the bias fields. We select an equal superposition ($\theta = \pi/2$) but
with a phase $\phi = \pi/4$. Here we see that the reachable set (in terms of $\Delta V_{x_0}$) is very much 
reduced compared to the previous example. Even though the fidelity is near one for a large range of 
bias voltages, the trace distance is significantly greater than zero until $t_{max} \simeq 100$ cycles and
$\Delta V_{x_0}\simeq 0.2\times 2e/C$. This is a result of the fact that the target state is not an energy
eigenstate and is therefore non-stationary for all bias values. The control is therefore much more sensitive
to small variations.
\begin{figure}
\begin{center}
\includegraphics[height=12cm]{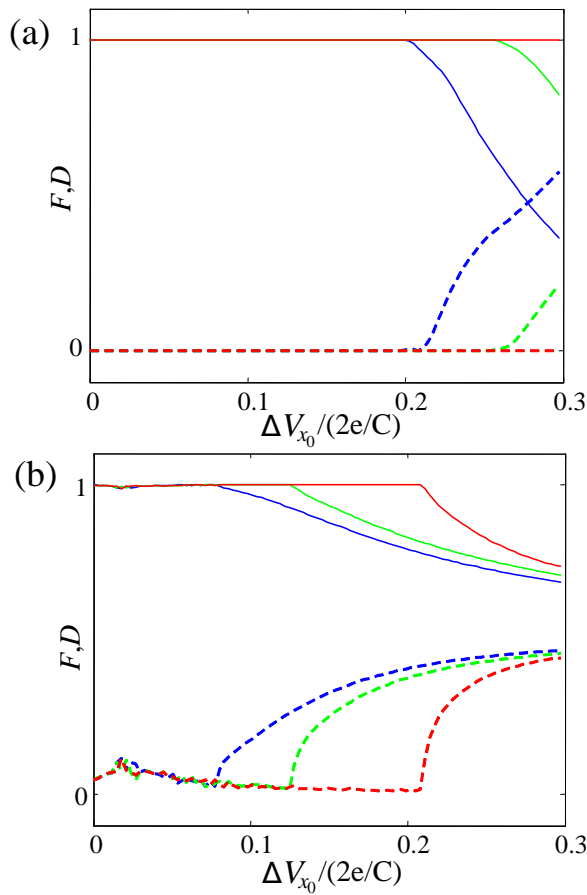}
\end{center}
\caption{(Color online) Fidelity ($F$ - solid lines) and trace distance ($D$ - dashed lines) achieved using
guidance algorithm as a function of the voltage bias value for  
different values of $t_{max}$: 20 qubit cycles (blue), 50 qubit cycles (green), and 100 qubit cycles (red); 
(a) target state is the excited state at nominal bias voltage, (b) target state is 
an arbitrary state with $\theta = \pi/2$ and $\phi = \pi/4$.}
\end{figure}

Figure 3 shows an example of the evolution of the qubit state under the proportional navigation
guidance and the control fields that were applied. The evolution starts in the ground state and
rapidly spirals around the Bloch sphere under the influence of the control, and gradually approaches the
excited state, spiralling in gradually as the controls applied reduce in size. This is a good
example of the benefit of this approach, where the controls subside to zero as the system approaches
the desired state. This has distinct experimental advantages because the bias fields will be static
immediately prior to $t_{go} = 0$, which means that the Hamiltonian is not varying rapidly when the
desired state is required. It is also noticeable that both control fields contain a dominant frequency
component that matches the coherent oscillation frequency of the qubit, indicating that a coherent
drive at the transition frequency is an important part of the control fields.
\begin{figure}
\begin{center}
\includegraphics[height=12cm]{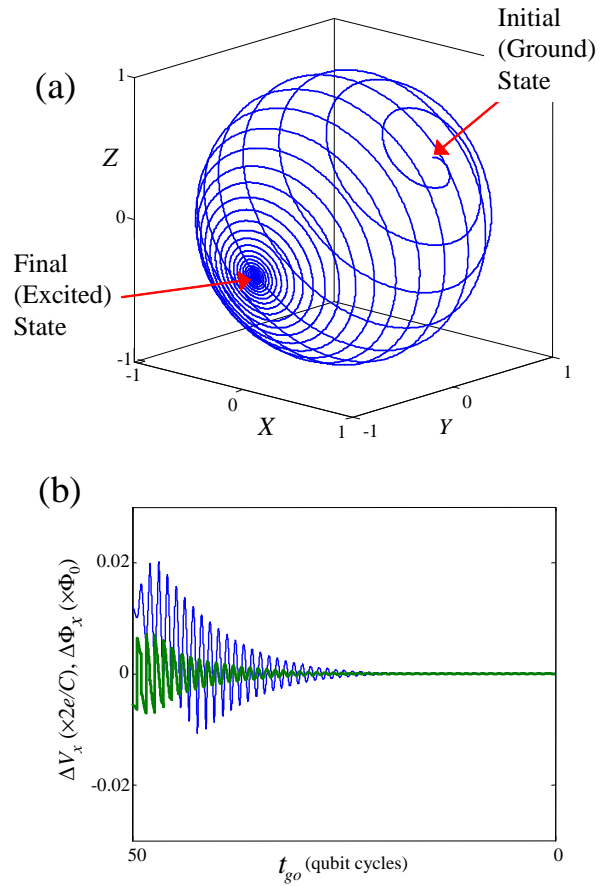}
\end{center}
\caption{(Color online) (a) Example trajectory of state on the Bloch sphere under proportional
guidance with the excited state as the target state and $\Delta V_{x_0}=0.1\times 2e/C$; 
(b) Control fields $\Delta V_{x_c}$ (green) and $\Delta\Phi_{x_c}$ (blue)
as a function of the $t_{go}$ for example trajectory in Fig 3(a).}
\end{figure}

\section{Imperfect Bias Fields}

\noindent
In classical guidance systems the guidance-control feedback loop must be robust enough
so that small perturbations from noise or imperfections in the control system 
are damped out and the system achieves its objective.
The main imperfections in classical guidance tend to come from uncertainties
in the physical parameters that define the transfer function between the accelerations
commanded by the guidance law and the actual accelerations achieved by the controls.
In the qubit guidance case, this is more difficult because of the problems
already mentioned in measuring the error signal and the algorithm discussed so
far is an open-loop control system. Instead, we require that the fidelity and trace 
distance of the final state be weakly sensitive to small variations in the experimental
control parameters. For example, the static (nominal) bias point may only be known to 
a certain accuracy. We should require that the final state is approximately correct
even if the static bias field is slightly off or if a small amount of dynamical 
noise is present in the fields. This is also of concern when the qubit is non-ideal, either
the gate capacitance is only known to a finite accuracy (so the scaling of the voltage biases
is inaccurate) or the Hamiltonian includes non-ideal terms (possibly due to variations in
the junctions coupling the island to the bulk material). In these cases the controls applied will
not necessarily generate rotations about axes exactly aligned to the the X- and Z-axes. 

In practice, we find that - in common with the performance of the guidance algorithm itself -
the sensitivity to noise is dependent on the target state. For static errors of the order
of $\Delta(\Delta V_{x_0})\simeq 10^{-5}\times 2e/C$ and $\Delta(\Phi_{x_0}) \simeq 10^{-4}\Phi_0$,
the performance of the algorithm for the example given in Fig 2(a) (i.e. the target state is the 
excited state) is very good. (The performance is also good for time-dependent errors, as long as the
cumulative errors during the control cycle are of the same order as these tolerances). 
Even for variations an order of magnitude larger than this, the 
performance is still acceptable for the excited state over a comparatively wide range of bias voltages, 
even if the reachable set is significantly reduced. The accuracy of the final state for the target state 
given in Figure 2(b) would be significantly less with errors of this size. The trace distance
between the final state and desired state is noticeably larger when bias errors are introduced.
For dynamical noise, we have also characterised the performance of the
algorithm in the presence of white noise (i.e. uncorrelated with a uniform frequency distribution), 
and the performance is similar to that for static bias
errors as long as the cumulative drift of the bias fields is less than the limits given above.

\section{First-Order Time Delays}

\noindent
In addition to noise, an experimental system is also likely to contain other imperfections.
The main one considered here is a constraint on the bandwidth allowed for the control
signal. We use a simple method of restricting the bandwidth in the control system by 
introducing a first order time-delay, which acts as a low-pass filter and has a transfer
function (Laplace transform) given by \cite{Ric79},
$$
F(s) = \frac{1}{1+sT_{d}}
$$
where $T_d$ is the time delay constant. An example of this type of delay is a low-pass
RC filter, which should be familiar from standard electrical circuit analysis. The effect of this
filter is to exponentially damp rapid variations in the controls, and in Figure 4 we show the effect of
such time delays on the performance of the guidance algorithm. Time delays of the order of $T_d \simeq 0.1-0.2$
cycles have little effect on the fidelity or trace distance of the final state for bias voltages
close to zero, but toward the far right (higher voltage biases), the excited state becomes harder to
reach, indicating that the bandwidth of the control signal is larger at these extreme values. By contrast,
the effect of first-order time delays on the case shown in Figure 2(b) would actually be less pronounced than
that shown in Figure 4 because the range of bias values from which the target state is reachable is already 
comparatively small.
\begin{figure}
\begin{center}
\includegraphics[height=6cm]{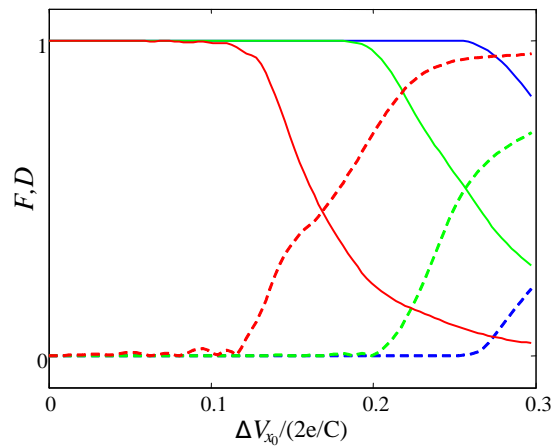}
\end{center}
\caption{(Color online) Fidelity ($F$ - solid lines) and trace distance ($D$ - dashed lines) achieved using
guidance algorithm as a function of the voltage bias value for different time delays: zero time delay 
(blue), 0.125 qubit cycle delay (green), and 0.25 qubit cycle delay (red); 
target state is the excited state at nominal bias voltage and $t_{max}= 50$ cycles.}
\end{figure}

\section{Closed-Loop Performance}

\noindent
In this section we consider the use of a simple measurement and Markovian feedback
mechanism to demonstrate how the open-loop guidance approach could be adapted using
existing quantum feedback techniques. As an example, we choose a simple model for
photon detection and (instantaneous) feedback. The model assumes that the qubit is weakly
coupled to a lossy reservoir and that projective measurements are made on this
reservoir. The results of the measurement are then used to modify the controls
applied to the qubit. This model may not be entirely realistic, because of 
problems with detecting single microwave photons and with the large bandwidths required
for a rapid feedback, but it demonstrates the general approach. The basic 
idea is to apply the guidance algorithm as described above
and to modify the control pulses, to allow for the reduced time to go, when a photon
is emitted and detected (detection is assumed to occur with efficiency $\eta$). Where
photons are not detected, the qubit will still be coupled to the lossy reservoir 
which will cause an additional (non-unitary) perturbation on the
otherwise coherent dynamics of the system and this is allowed for in the modelling but
not in the guidance-control algorithm.

The measurement mechanism is modelled using a
{\it quantum trajectory} approach \cite{Car93,Wis96}, corresponding to an
unravelling of the Markovian Master equation for the qubit reduced
density operator (after performing a partial trace over the lossy reservoir). 
In this paper, we choose the {\it quantum jumps}
approach \cite{Car93,Wis96}, which is suitable for modelling spontaneous
emission processes and is computationally efficient \cite{Wis99}.
Physically, this unravelling corresponds to the detection or absorption
of spontaneously emitted photons on a time scale that is significantly
faster than any of the time scales present in the quantum system.
All unravellings reproduce the Master equation evolution when averaged, and
the individual quantum `trajectories' for the qubit are described by a model given in
reference \cite{Wis99}. 

The spontaneous emission process and subsequent detection of the photon
introduces quantum jumps that project into the instantaneous ground state
of the qubit. The probability of a spontaneous decay occurring during a small 
- but finite - time interval, $\delta t$, is
$\gamma\left<\hat{c}^{\dagger}\hat{c}\right>\delta t$, where $\hat{c}^{\dagger}$ and $\hat{c}$ are
the raising and lowering operators for the (instantaneous) qubit energy states
respectively. During each time interval where no spontaneous decay occurs, a non-unitary evolution operator
$$\hat{\Omega}_{0}(dt)=1-\frac{i}{\hbar}\hat{H}_{qu}\delta t-\frac{\gamma}{2}\hat{c}^{\dagger}\hat{c} \delta t$$
is applied to the qubit state. When a decay occurs, an operator
$$\hat{\Omega}_{1}(dt)=\sqrt{\gamma \delta t}\:\hat{c}$$ 
is applied to the qubit state. Each run of the simulation
produces a subjective `trajectory'. For each trajectory, feedback control is invoked 
if a photon is detected, with probability 
$\eta$, so that the state evolution is conditional upon the detections and then averaged
over multiple realisations. The result is averaged over many runs
to provide an estimate of the mixed state density matrix
for the qubit, from which we can calculate purity, fidelity and trace distance.
(Some noise is still present in the average density matrix and the performance measures, 
but this is relatively small and is due to the limited number of runs (typically 500-1000)
which is dictated by computational constraints). 
\begin{figure}
\begin{center}
\includegraphics[height=12cm]{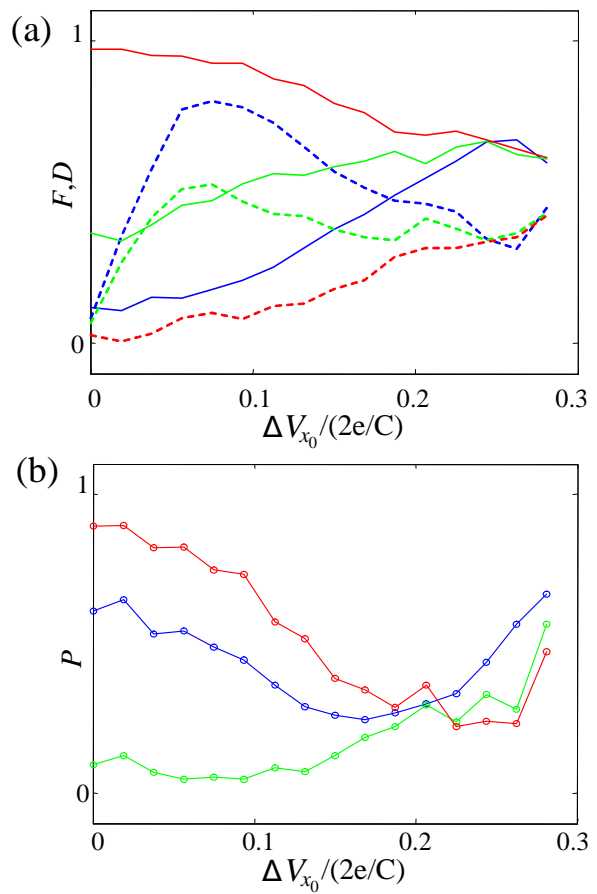}
\end{center}
\caption{(Color online) (a) Fidelity ($F$ - solid lines) and trace distance ($D$ - dashed lines), and (b) Purity ($p$ - circles) 
achieved using guidance algorithm as a function of the voltage bias value for $\gamma = 0.05$/cycle and 
$\eta = 0.0$ (no feedback - blue), $\eta = 0.5$ (green), and $\eta = 1.0$ (red); 
target state is the excited state at nominal bias voltage and $t_{max}= 50$ cycles.}
\end{figure}

Figure 5 shows an example of the fidelity, trace distance and purity for this simple closed-loop control 
system, corresponding to one of the examples shown in Figure 2(a) with the damping rate $\gamma = 0.05$/qubit 
cycle. If the damping rate were significantly smaller than this, the probability of a qubit undergoing
a spontaneous decay during the period of control would be negligible and the system would reduce to the
open-loop case already discussed. If the damping rate were significantly higher than this, guidance-control (open-loop
or closed-loop) would be impossible. For the case shown, the probability of undergoing a transition during the
control period is significant. Without feedback (or very inefficient photon detection $\eta=0$ - blue lines) 
the fidelity and trace distance are very low for small voltage bias values, and the purity is fairly high.
This is an indication that most of the qubits will spontaneously emit photons and decay back to the ground
state. The main controls are applied near the start of the control period (see Figure 3(b)), which will
tend to leave the qubit near the ground state once a photon has been emitted. In fact, the two properties
are related, since the qubit is only likely to emit a photon once the control has brought the system
close to the excited state in the first place. As the bias voltages are increased, the excited state is harder
to reach (and therefore occurs later in the control cycle) and the emission probability consequently goes down.
This is the cause of the minimum in the purity for the $\eta = 0$ case, where the two effects balance out,
so that the mid-range voltages are more likely to be mixed between the ground and excited states. 

Where the detection probability is non-zero, the green and red curves in Figure 5, feedback is allowed and the 
controls can be modified when a photon is detected. In the case of $\eta = 0.5$ the purity is significantly
reduced because there is a chance of being near either state for most values of the bias voltage, either through
decay and detection, decay and non-detection, and non-decay. However, as desired, the fidelity increases
as the detection probability increases, indicating that the feedback is working correctly. Even so, even
with $\eta = 1$, the closed loop performance does not reach the open-loop, non-dissipative performance. 
This is because the spontaneous emission reduces the effective time available for the control, and multiple
jumps are likely to occur for some bias voltages. The average number of jumps is dependent on $\eta$ and 
$\Delta V_{x_{0}}$, but for $\eta = 1$ and $\Delta V_{x_{0}}\simeq 0$ two or more jumps are not uncommon.


\section{Discussion}

\noindent
In this paper we have presented a generalisation of a classical guidance law to the problem of
control of a qubit state on the Bloch sphere. We have chosen the proportional navigation guidance law because of
its relative simplicity and its resultant widespread use in classical guidance and control 
systems. We have demonstrated that this guidance law can be used to generate 
an arbitrary quantum state from the ground state of a superconducting charge qubit using the standard 
control fields (voltage and magnetic flux bias). The controls produced by this guidance law
are relatively robust to imperfections in the control fields and to first order time delays,
implying that the control signals have a comparatively low bandwidth. This should
make it possible to control the state of the qubit using signals fed through low-pass 
transmission lines. We have also suggested a simple method to allow the generalised guidance law
to be included in a closed-loop (Markovian) control system.

Although the ability to control the state of a single qubit with a high degree of accuracy is important
for possible quantum information processing device, the ability to control the collective behaviour
of multiple qubits is also of great interest. The ability to visualise the control of the qubit state
on the Bloch sphere is useful in understanding the guidance mechanism, but it is not essential. 
Generalising the guidance algorithm to higher dimensional settings (multiple qubits or N-level systems)
simply requires an understanding of the group structure of the space and the ability to create appropriate
control Hamiltonians from the generators of the group \cite{Sch04}. As long as the group generators
(or the restricted set of generators available to the control system) allow the state space to be explored
fully, then it should be possible to generalise the guidance algorithm described in this paper to higher
dimensional systems.


\end{document}